\documentclass[prd,twocolumn,nofootinbib]{revtex4}
\usepackage{graphicx, epsfig}
\usepackage{color}
\usepackage{mathrsfs}
\usepackage {amssymb}
\usepackage {amsmath}
\usepackage{soul}

\newcommand{\nc}{\newcommand}

\nc{\half}{\frac{1}{2}}
\def\M{M_{\rm Pl}}

\def\epsc{\epsilon_{{\rm c}}}

\def\kpivot{k_{{p}}}

\newcommand{\refeq}[1]{(\ref{#1})}

\input{epsf.sty}

\begin{document}

\title{Geometrical Destabilization of Inflation}

\author{S\'ebastien Renaux-Petel$^{1,2}$ and Krzysztof Turzy\'nski$^3$}

\affiliation{$^1$Institut d'Astrophysique de Paris, UMR-7095 du CNRS, Universit\'e Pierre et Marie Curie, 98~bis~bd~Arago, 75014 Paris, France,}
\affiliation{$^2$Sorbonne Universit\'es, Institut Lagrange de Paris, 98 bis bd Arago, 75014 Paris, France}
\affiliation{$^3$Institute of Theoretical Physics, Faculty of Physics, University of Warsaw, Pasteura 5, \mbox{02-093 Warsaw}, Poland.}

\date{\today}

\begin{abstract}
We show the existence of a general mechanism by which heavy scalar fields can be destabilized during inflation, relying on the fact that the curvature of the field space manifold can dominate the stabilizing force from the potential and destabilize inflationary trajectories. We describe a simple and rather universal setup in which higher-order operators suppressed by a large energy scale trigger this instability. This phenomenon can prematurely end inflation, thereby leading to important observational consequences and sometimes excluding models that would otherwise perfectly fit the data. More generally, it modifies the interpretation of cosmological constraints in terms of fundamental physics. We also explain how the geometrical destabilization can lead to powerful selection criteria on the field space curvature of inflationary models.
\end{abstract}

\maketitle

{\bf Introduction.}---Recent cosmic microwave background data from the \textit{Planck} and \textit{BICEP2/Keck} collaborations \cite{Ade:2015lrj,Ade:2015tva} constrain the tensor-to-scalar ratio $r < 0.12$ ($95 \%\,\, {\rm C.L.}$)
and  the spectral index of primordial density perturbations  $n_s=0.968 \pm 0.006$ ($68 \%\,\, {\rm C.L.}$). 
The simplest slow-roll single-field inflationary models are in perfect agreement with these data and the lack of measurable primordial non-Gaussianities \cite{Ade:2013ydc,Ade:2015ava}. Amongst them, models coming with a concave plateau potential, such as Starobinsky inflation \cite{Starobinsky:1980te} and its numerous variants (see, e.g., \cite{Kallosh:2013xya,Ellis:2013xoa,Carrasco:2015rva}), are observationally favored. Despite this phenomenological success, embedding inflation into a realistic high-energy context remains a highly nontrivial task (see Refs. \cite{Linde:2005ht,Baumann:2014nda} for reviews),  
as inflation is an ultraviolet-sensitive phenomenon. 
One manifestation of these difficulties is the so-called $\eta$ problem, i.e.,
the fact that even Planck-suppressed corrections to an otherwise flat enough potential generically ruin inflation. 
Another challenge is the ubiquitous presence of extra scalar fields in models constructed in supergravity or string theory. In general, these fields participate in the inflationary dynamics and can substantially modify the corresponding observable predictions. In this respect, the simplest theoretical hope is to stabilize these extra scalars by providing them with a large mass, exceeding the value of the Hubble parameter $H$ during inflation.

In this Letter, we show the existence of a very general geometrical mechanism by which the noninflationary 
degrees of freedom can be destabilized {\it during} inflation, even if they have large masses in the static vacuum. This mechanism relies on the fact that generic multifield inflationary models are nonlinear sigma models, i.e.,
the kinetic part of their action reads \mbox{${\cal L}_{\rm kin}=-\frac{1}{2} G_{IJ}(\phi^K) \partial_{\mu} \phi^I \partial^{\mu} \phi^J$}, where the manifold described by the field space metric $G_{IJ}$ is generally curved. It is well known that the curvature of space manifests itself in the geodesic deviation:
\begin{equation}
\frac{D^2 \xi^{\mu}}{D \tau^2}=R^{\mu}_{\,\,\alpha \beta \nu} V^{\alpha} V^{\beta} \xi^{\nu}\,,
\label{deviation}
\end{equation}
indicating how two neighboring geodesics separated by $\xi^{\mu}$ tend to fall closer to or apart from each other, and where  $D/D \tau$ denotes the covariant derivative along the tangent vector $V^{\mu}$. Although many authors noted the possible impact of the field space curvature on the inflationary dynamics (see, e.g., \cite{Sasaki:1995aw,multi1,Cremonini:2010ua,multi2}), 
we argue that this physical effect can have much further reaching consequences than hitherto noticed. To give a simple and concrete example: in two-field models (in which the curvature tensor can be solely expressed in terms of the field space metric and the Ricci scalar $R_{{\rm fs}}$) with hyperbolic geometry, i.e., with negative 
$R_{{\rm fs}}$, all inflationary trajectories, even those aligned with field space geodesics, tend to become unstable. 
This phenomenon can prematurely end inflation and modify its predictions in a way that can be very substantial, sometimes excluding models that would otherwise perfectly fit the data.
In particular, the stabilization of noninflationary degrees of freedom
 with a steep potential easily fails to counterbalance the geometrical destabilization 
mentioned above. Effectively, the resulting destabilization of the inflationary trajectory is akin to the one occurring in hybrid inflation \cite{Linde:1993cn}, but the scope of our mechanism is much broader, as it potentially affects
 all inflationary models. We stress that generic high-energy corrections suppressed by a large energy scale can trigger 
such a destabilization, and we provide a powerful selection criterion on the field space curvature of inflationary models. Our mechanism is therefore of utmost importance for the confrontation of high-energy physics with cosmological observations.

{\bf General mechanism.}---Generic models 
of inflation in supergravity come with a curved field space, inherited from the K\"ahler potential of the theory. Naturalness considerations within such a setup 
lead to quite a stringent upper bound on at least one projection of the curvature tensor \cite{sugrainf}.
A precise value of this bound depends on particular assumptions,
but it can be concluded that in two-field models having
a negative field space curvature is the most straightforward way
for embedding inflation in supergravity. For instance, recent popular two-field supergravity realizations of
$\alpha$-attractor models are characterized by a negative $R_{{\rm fs}} =-2/(3\alpha  \M^2)$ that is 
large in the interesting limit of small $\alpha$ \cite{Carrasco:2015rva}. Also, from the effective field theory viewpoint, it can be expected that the low energy effective Lagrangian describing an inflaton $\phi$ coupled to an extra scalar $\chi$ contains an operator $-(\partial\phi)^2\chi^2/M^2$, giving negative contributions to $R_{{\rm fs}}$
of the order ${\cal O}(1/M^2)$, where $M \gg H$ is the scale of new physics beyond the single-field description.
Motivated by these examples, we study
in the following the phenomenological consequences of inflation with field space curvature $\sim 1/M^2$ characterized by mass scales lying between the Hubble and the Planck scales.

The linearized evolution equations of fluctuations in nonlinear
sigma models have been known for more than two decades and read \cite{Sasaki:1995aw}:
\begin{eqnarray}
{\cal D}_t {\cal D}_t Q^I  +3H {\cal D}_t Q^I +\frac{k^2}{a^2} Q^I +M^I_{\,J} Q^J=0\,.
\label{pert}
\end{eqnarray}
Here, $Q^I=\phi^I(t,\boldsymbol{x})-{\bar \phi}^I(t)$ are the field fluctuations (in the spatially flat gauge) above their background values, and ${\cal D}_t A^I \equiv \dot{A^I} + \Gamma^I_{JK} \dot \phi^J A^K$ is a covariant time derivative in field space. The crucial physical information lies in the effective mass matrix 
\begin{eqnarray} 
\label{masssquared}
M^{I}_{\,J} &=& V^{I}_{; J} - \mathcal{R}^{I}_{\,KLJ}\dot \phi^K \dot \phi^L -\frac{1}{a^3}\mathcal{D}_t\left(\frac{a^3}{H} \dot \phi^
I \dot \phi_J\right)\,.
\end{eqnarray}
Besides the expected Hessian of the potential, 
the second term, analogous to the one in the geodesic deviation equation \refeq{deviation}, drives the mechanism we mentioned above 
(the third one represents well-known kinematical effects). For simplicity, we concentrate on the case of two fields
in the following. To study the stability of the inflationary trajectory, we project Eq.~\refeq{pert} into the direction perpendicular to the background velocity, obtaining the 
super-Hubble
evolution equation
$\ddot Q_s+3 H \dot Q_s +m^{2}_{s {\rm (eff)}} Q_s= 0$ 
for the so-called entropic perturbation $Q_s \equiv e_{s I} Q^I$. 
Here, the effective mass squared in Hubble units reads
\begin{equation}
\frac{m^{2}_{s {\rm (eff)}}}{H^2} = \frac{V_{;ss}}{H^2} +3 \eta_\perp^2+ \epsilon  \, R_{{\rm fs}} \M^2\,,
\label{ms2}
\end{equation}
where $V_{;ss}=e_s^I e_s^J(V_{,IJ}-\Gamma_{IJ}^K V_{,K})$ is the projection of the covariant second derivative of the potential along the entropic direction $e_s^I$, and the dimensionless
parameter $\eta_{\perp}$ measures the rate at which the trajectory deviates from a field space geodesic \cite{Gordon:2000hv,multi1}.When $R_{{\rm fs}}$ is negative, 
which, as we have seen, is not a restrictive condition,
it acts in the direction of rendering the effective entropic mass tachyonic. As $\epsilon$ is positive and usually grows during inflation, reaching $1$ at the end of inflation, it is crucial to take into account this universal geometrical contribution, even in the case of a large ``static mass'' $V_{;ss}$. For instance, with
values $M= {\cal O}(10^{-2},10^{-3}) \M$, like the typical string scale or Kaluza-Klein scale in string theory contexts, or the scale of grand unification, and with a stabilization value $V_{;ss} \sim 100 \,H^2$---already much larger than in most examples in the literature---the entropic mass becomes tachyonic when $\epsilon$ reaches a critical value $\epsc$ as low as $10^{-4}$ or $10^{-2}$ (assuming geodesic motion for simplicity). 
Like in hybrid inflation, the exponential growth of entropic perturbations caused by this tachyonic (spinodal) instability affects all Fourier modes that cross the Hubble radius before its onset. The subsequent cosmological evolution depends on the backreaction of these fluctuations on the inflationary trajectory, with theoretical uncertainties and model-dependences similar to the ones in hybrid inflation. However, even in the simplest conservative approach, in which inflation ends abruptly and entropic fluctuations do not have the time to affect the curvature perturbation, observational consequences do exist and are important: with such a premature end of inflation, the observable scales exit the Hubble radius on a typically flatter part of the potential, which tends to bring $n_s-1$ and $r$ closer to zero.

{\bf A minimal realization.}---We now 
describe a simple and rather universal setup that exhibits 
the geometrical destabilization depicted above. Let us start with a model of slow-roll inflation, with $\phi$ playing the role of the inflaton and with the Lagrangian 
$\mathcal{L}_{\phi}=- \half (\partial\phi)^2 -V(\phi)$. In realistic models embedded in high-energy physics, there exist extra scalar fields beyond the inflaton and the simplest expectation is that they are heavy during inflation (with masses much larger than $H$), and that their configuration corresponds to the minimum of the potential in directions orthogonal to the inflationary one. We model this by adding a field $\chi$ described by a simple Lagrangian $\mathcal{L}_{\chi}=- \half (\partial\chi)^2 -\half m_h^2\chi^2$ with $m_h^2 \gg H^2$, so that $\chi$ is anchored at the bottom of the inflationary valley at $\chi=0$. 
We also assume that the fields $\phi$ and $\chi$ interact via a higher-order operator $-(\partial\phi)^2\chi^2/M^2$, as may generally be expected from the effective theory point of view, 
and where $M$ denotes the associated energy scale of new physics. The absence of operators linear in $\chi$ is required for consistently having $\chi=0$ solve the equations of motion, 
whereas higher powers of $\chi/M$ are suppressed near $\chi=0$. This very simple setup provides a realization of the geometrical destabilization: along the inflationary valley $\chi=0$,
the Ricci scalar $R_{{\rm fs}}\simeq -\frac{4}{M^2}$
resulting from the modified kinetic terms
is negative and can destabilize the trajectory, 
as the effective entropic mass squared reads 
\begin{equation}
m^{2}_{s {\rm (eff)}} = m_h^2 -4\, \epsilon(t) H^2(t)  \left( \frac{\M}{M}\right)^2\,.
\label{ms2simple}
\end{equation}
If the kinetic energy density $\epsilon H^2 \M^2=\half \dot \phi^2$ grows during inflation, $m^{2}_{s {\rm (eff)}}$ can turn from positive to negative, triggering the instability. This necessary condition for the destabilization reads $\epsilon_2 \equiv \dot \epsilon/(H \epsilon)>2 \epsilon$,
and simply states that for the effective mass in Eq.~\refeq{ms2} to become tachyonic, the rate of change of $\epsilon$ should be large enough to compensate the increase of $m_h^2/H^2$ during inflation. In the slow-roll regime, this translates in terms of the potential as $2 V''V < {V'}^2$. This condition is not very restrictive: it reads $p<2$ for monomial potentials $V \propto \phi^p$, and, more generally, it shows that the effect described here applies to the observationally preferred models with concave potentials. Eventually, let us stress that our effective Lagrangian 
\begin{equation}
 {\cal L}=-  \left(1+ \frac{2\chi^2}{M^2}\right) \frac{(\partial \phi)^2}{2}
 -V(\phi)-\frac{(\partial \chi)^2}{2}-\frac{m_h^2 \chi^2}{2}
\label{simple-realization}
\end{equation}
corresponds to the local expansion 
along the inflationary valley 
of many concrete models, including no-scale realizations of Starobinsky inflation in supergravity and models of $\alpha$-attractors (see, e.g., Refs. \cite{Ellis:2013xoa,Carrasco:2015rva} for recent examples), 
in which the geometrical destabilization is overlooked.

{\bf Similarity and differences with hybrid inflation.}---Our geometrical mechanism and hybrid inflation share an essential feature: entropic fluctuations grow exponentially due to a tachyonic instability, which destabilizes the background.
However, the two setups differ both by the physical mechanism causing this instability and the context in which it takes place. In hybrid inflation, the potential for the ``waterfall'' field $\chi$ and its coupling to the inflaton $\phi$ is chosen to end inflation. It determines both the location in field space at which the instability kicks in and the stable equilibrium point of the system away from $\chi=0$. However, before that point is reached, one may wonder whether the tachyonic growth of the entropic perturbations terminates inflation abruptly, or there may be a second phase of inflation driven by the waterfall field, and most importantly how inflationary observables are affected by these phenomena.
Addressing these questions requires working beyond standard linear perturbation theory in a stochastic framework that takes into account the backreaction of short-wavelength perturbations on the long-wavelength modes, together with the evolution of perturbations on the associated modified background. The exponential  growth of perturbations may also lead to the formation of primordial black holes \cite{GarciaBellido:1996qt}. Eventually, in the case of a $\mathbb{Z}_2$ symmetry $ \chi \to - \chi$, manifest in the minimal realization discussed above, nonperturbative 
calculations and lattice simulations are required to tackle spontaneous symmetry breaking (i.e., tachyonic preheating) 
and the creation of inflating topological defects \cite{Felder:2000hj}.
These various theoretical challenges are still the subject of an intense activity more than two decades after the birth of hybrid inflation, and the answers are both model and parameter dependent, 
if not under debate (see, e.g., Ref.~\cite{Mulryne:2009ci}). 
The same questions apply to the geometrical destabilization that we discuss here, and providing definite answers is beyond the scope of this Letter, all the more as our setup exhibits notable differences compared to hybrid inflation. First of all, the inflaton and the additional field $\chi$ are kinetically coupled, and the destabilization does not arise at a critical value of the inflaton but rather of its velocity. In addition, while our simplifying assumptions---the quadratic potential for $\chi$ and the form of the kinetic coupling---can be justified as leading terms in an expansion around $\chi=0$, the growth of $\chi$ fluctuations can be so fast that this description ceases to be reliable, and the knowledge of the full Lagrangian beyond our effective model becomes necessary. In short, determining the fate of the inflationary phase and the impact on cosmological observables requires substantial theoretical developments and depends on the completion of the model away from the inflationary valley.

{\bf A prototypical example.}---Let us 
consider a concrete prototypical example: Starobinsky inflation, with a potential 
$V(\phi)=\Lambda^4 [1-\mathrm{exp}(-\sqrt{2/3}\, \phi/\M)]^2$. In the single-field picture, it generates a curvature power spectrum 
${\cal P}_{\zeta}(k)=H^2_\star/(8 \pi^2\M^2 \epsilon_{\star})$, a spectral index $n_s-1=-2 \epsilon_\star-\epsilon_{2 \star}$, and a 
tensor-to-scalar ratio $r=16 \epsilon_\star$, where $\star$ denotes evaluation at the Hubble crossing time such that $k=a H$. As usual, we can adjust the inflationary scale $\Lambda$ so that ${\cal P}_\zeta(\kpivot)$ agrees with the observed amplitude $A_s \simeq 2.2 \times 10^{-9}$ at the pivot scale $\kpivot=0.05 \, {\rm Mpc^{-1}}$. The identification of $\Delta N_{p}$, 
the number of e-folds between the Hubble exit 
for the pivot scale and the end of inflation, 
depends on the details of reheating and should be self-consistently determined (see, e.g., Ref.~\cite{Martin:2014nya}). 
In the following, we take the representative central value $\Delta N_{p}=55$. 
When inflation ends by slow-roll violation around $\epsilon=1$, this generates the well-known values $n_s=0.965$ and $r=0.0036$, in perfect agreement with the data.

In the presence of the tachyonic instability described here, a precise treatment of the inflationary perturbations is subject to a considerable theoretical uncertainty. We address this uncertainty by comparing the following two simple and rather conservative approaches. In the first one,
we assume that inflation ends abruptly after the critical point and entropic fluctuations do not have time to affect the curvature perturbation on cosmological scales. Within this approximation, it is straightforward to derive modified observational predictions for any inflationary model, once values are given for the ``stabilizing'' 
mass $m_h$ and the energy scale $M$ associated with
the curvature of the field space. For the simplest models currently in agreement with the observational data, the inflationary potentials become flatter as one goes backward in time during inflation (corresponding to smaller $\epsilon$ and $\epsilon_2$). The general trend is therefore to move the predictions closer to scale invariance and towards a smaller amplitude of gravitational waves, with important consequences for assessing the status of inflationary models \cite{preparation}.

We compare these results with predictions of the second approach, in which at the critical point we shift the classical background value of $\chi$ away from zero to a typical value, which we take to be $H_{{\rm c}}/2 \pi$, and then follow the evolution of the coupled two-field system. After a rapid growth of $\chi$, the system continues evolving along a stable trajectory with slowly evolving $\chi\sim \mathcal{O}(M)$, i.e., a second phase of inflation begins. Thanks to the slowing down of the inflaton field by a noncanonical normalization, this phase can last very long: in the numerical examples presented here its duration ranges from a few to over a thousand e-folds. Interestingly, this phase corresponds to a large value of $\eta_\perp$ and can be in certain situations described by an effective one-field model with a modified dispersion relation \cite{Tolley:2009fg,Baumann:2011su} or a model with transient tachyonic instability at the Hubble radius crossing \cite{Cremonini:2010ua}. Inflation then ends in a standard fashion by slow-roll violation, $\epsilon=1$, before the fields rapidly converge to the stable potential minimum at $(\phi,\chi)=(0,0)$.

\begin{figure}
 \hspace{-0.8cm}
  \includegraphics*[width=6.6cm]{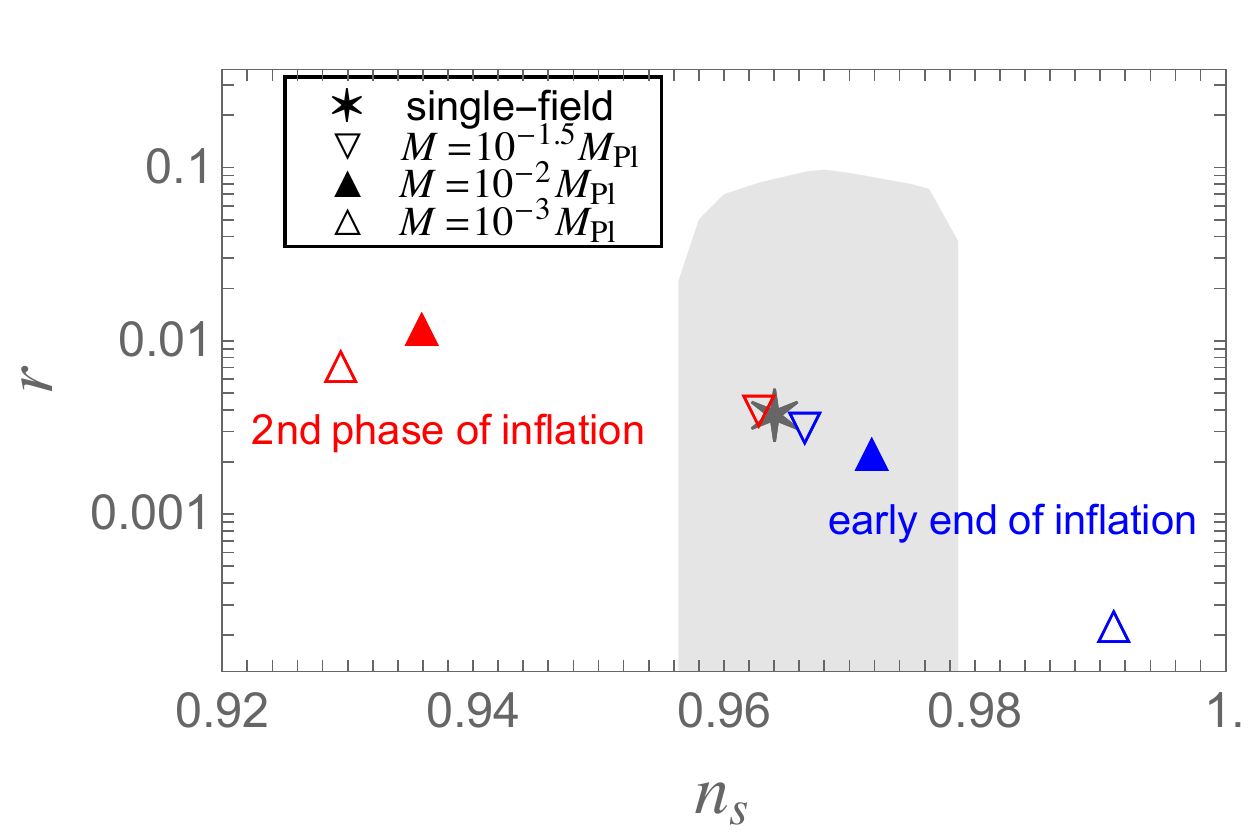}
\caption{Modified predictions for the spectral index and tensor-to-scalar ratio
in our prototypical example based on the Starobinsky potential, for the scale that 
crosses the Hubble radius 55 e-folds before the end of inflation. The shaded area corresponds to the {\em Planck} $95\%\,\mathrm{C.L.}$ constraints \cite{Ade:2015lrj}.   \label{ns-r}}
 \end{figure}

In Fig.~\ref{ns-r}, we show the predictions in the $(n_s,r)$ plane for our prototypical example: modified Starobinsky inflation with $m_h^2=10^2H_{{\rm c}}^2$ at the onset of the tachyonic instability, with $M/\M=10^{-1.5}$, $10^{-2}$, and $10^{-3}$,  and 
in both approaches to the evolution of the system through the instability.
It may well be that inflation ends during the phase of expulsion of $\chi$ away from the unstable point, but we nonetheless think that studying the kind of second phase of inflation described above is interesting for different choices of the field space metric
and potential, as we numerically observed that it corresponds to an attractor point.

{\bf A universal bound on the field space curvature.}--- Let us consider a generic two-field nonlinear 
sigma model, in which a geometrical destabilization arises due to a negative field space curvature, with a transition from a positive to a negative super-Hubble entropic mass squared \refeq{ms2}. Let us assume that the instability abruptly terminates inflation, introduce the mass scale $M_\star$ such that $| R_{{\rm fs} \,\star} |=4/M_\star^2$, and simply write down that $m^{2}_{s {\rm (eff)}}$ is positive at the time of Hubble crossing for the pivot scale, i.e.,
$\left( \epsilon | R_{{\rm fs}} | \M^2 H^2 \right)_\star   < M_h^2 \equiv  \left( V_{;ss}+3 \eta_\perp^2 H^2 \right)_\star $. Since the power spectrum is always greater than the single-field result, we readily obtain
\begin{equation}
\frac{M_\star}{H_\star} > \frac{1}{\sqrt{2 \pi^2 A_s}} \left( \frac{H_\star}{M_h} \right) \simeq 5000 \left( \frac{H_\star}{M_h} \right)\,,
\label{lowerbound}
\end{equation}
where we used the cosmic microwave background normalization. 
Models with a smaller value of $M_\star$ are excluded.
In our minimal realization, the quantities $M_\star$ and $M_h$ simply reduce to $M$ and $m_h$; in that case, for $M> 2 \M H_{{\rm end}}/m_h$, the geometrical destabilization does not arise and the inflationary phase is effectively single-field, while for intermediate values of $M$, inflation ends prematurely. The bound \refeq{lowerbound} gives a powerful model-independent selection criterion on inflationary models and the geometry of their field space. 
Unlike the constraints based on the (non)observation of non-Gaussianities, it 
enables one to constrain interactions at energies well above $H$ using 
only two-point correlation functions.
In our era of primordial cosmology characterized by data consistent with many simple models, we believe that such theoretical guidance is of utmost importance.

{\bf Perspectives and generalizations.}---As we mentioned, 
supergravity models of $\alpha$-attractors \cite{Carrasco:2015rva} are locally described by the effective Lagrangian \refeq{simple-realization} along the inflationary valley. These models are characterized by a scalar curvature $R_{{\rm fs}} \M^2=-2/(3\alpha)$ that is large for small $\alpha$. However, these models also have the built-in feature that $\epsilon \propto \alpha$ during the bulk of inflation. As a result, one can check that our geometrical destabilization can only arise at the end of inflation, when $\epsilon$ approaches $1$. The same holds true in several realizations of Starobinsky inflation in supergravity \cite{Ellis:2013xoa}, there because $R_{{\rm fs}} \M^2={\cal O}(1)$. We hypothesize that the geometrical instability can impact the phase of reheating in these models, and more generally in models with $\epsc \sim 1$.

In $N$-field inflationary models, the same threat of tachyonic instabilities is present in the full $(N-1)$-dimensional 
entropic sector, where the role of the Ricci scalar is replaced by the relevant projections of the Riemann curvature tensor. We stress therefore that the geometrical destabilization is not limited to hyperbolic two-field models on which we concentrated for simplicity.

We discussed here destabilization of very heavy fields, $m_h \gg H$, but it 
would be equally interesting to study the even more dramatic impact of our mechanism in models where the extra scalars have intermediate masses of the order the Hubble parameter (see, e.g., Refs.~\cite{Chen:2009we,Baumann:2011nk,McAllister:2012am}). 
We also concentrated on models in which $\epsilon$ steadily increases during inflation, but features in the potential can lead to transient slow-roll violations, and therefore temporarily large values of $\epsilon$ triggering the instability. Eventually, it would be interesting to relate the energy scale of new physics in our minimal realization to constraints 
on primordial non-Gaussianities, and to take into account our geometrical effect in studies of the inflationary landscape, which mainly concentrate on the structure of the potential, but for simplicity with a trivial field space metric (see, e.g., Ref.~\cite{Frazer:2011br}).

Let us eventually note that our mechanism is reminiscent of the well-known $\eta$ problem: in each case, higher-order operators suppressed by a large energy scale, even $\M$, can substantially modify the inflationary dynamics, ruining the required flatness of the inflationary direction in the case of the $\eta$ problem, and the required large curvature of the orthogonal directions in the situation described here.

{\bf Conclusions.}---We have 
described a very general mechanism of destabilization of scalar fields during inflation. It relies on the fact that the curvature of the field space can render inflationary trajectories unstable, dominating the stabilizing force from the potential. We described a simple and rather universal setup in which irrelevant operators suppressed by a high energy scale generate this effect. The resulting phenomenology varies depending on the details of the dynamics after the onset of the instability; in all cases, our mechanism can have important observational consequences, sometimes excluding models that would otherwise perfectly fit the data. We hope this mechanism, offering a new way to end inflation and unveiling another aspect of the UV sensitivity of inflation, will enable one to further use cosmological observations as a laboratory to probe physics at the highest energy scales.

\textit{Acknowledgments.} We are grateful to J\'er\^ome Martin, Shinji Mukohyama, and Ivonne Zavala, and particularly to David 
Mulryne and Vincent Vennin,  for very useful discussions. We would also like to thank the anonymous referees who helped us in improving the content and the clarity of our Letter. 
The work of S.R-P was supported by French state funds managed by the ANR within the Investissements
d'Avenir programme under reference ANR-11-IDEX-0004-02. K.T. is partly supported by Grant No. 2014/14/E/ST9/00152 from the National Science Centre, Poland.

%%%%%%%%%%%%%%%%%%%%%%%%%%%%%%%%%%%%%%%%%%%%%%%%%%%%%%%%%%%%%%%%%%%%%%%%%%%%%%%%%%%%%%%%%%%%%%%%%%%%%%
%\section*{References}

\end{document}